\documentclass[5p,times]{elsarticle}
\usepackage{ecrc}
\makeatletter
\let\origpprintTitle\ps@pprintTitle
\def\ps@pprintTitle{%
  \origpprintTitle
  \def\@oddfoot{\parbox{\columnwidth}{\footnotesize\itshape
    2213-8463\/ \copyright\ 2026 The Authors. Published by Elsevier Ltd.
    This is an open access article under the CC BY-NC-ND license
    (\url{http://creativecommons.org/licenses/by-nc-nd/4.0/})\\
    Peer-review under responsibility of the scientific committee of the NAMRI/SME.}}%
  \let\@evenfoot\@oddfoot}
\makeatother
\usepackage{enumitem}
\usepackage{amsmath}
\usepackage{mathtools}
\usepackage{booktabs}
\usepackage{algorithm}
\usepackage{algorithmic}
\usepackage{caption}
\usepackage{mdframed}
\usepackage{xcolor}
\usepackage{multirow}

\AtBeginDocument{
  \setlength{\abovedisplayskip}{5pt}
  \setlength{\belowdisplayskip}{5pt}
  \setlength{\abovedisplayshortskip}{1pt}
 
}

\volume{00}

\firstpage{1}

\journalname{Manufacturing Letters}

\runauth{Author name}

\jid{MFGLET}

\usepackage{amssymb}
\usepackage[figuresright]{rotating}

\usepackage[bookmarks=false]{hyperref}
    \hypersetup{colorlinks,
      linkcolor=blue,
      citecolor=blue,
      urlcolor=blue}

\begin{document}
\begin{frontmatter}



\dochead{54th SME North American Manufacturing Research Conference (NAMRC 54, 2026)}%

\title{An LLM-Assisted Multi-Agent Control Framework for Roll-to-Roll Manufacturing Systems}


\author{Jiachen Li} 
\author{Shihao Li}
\author{Christopher Martin}
\author{Zijun Chen}
\author{Dongmei Chen}
\author{Wei Li\corref{cor1}}
\ead{weiwli@austin.utexas.edu}

\address{Walker Department of Mechanical Engineering,
University of Texas at Austin, Austin, TX 78712}

\begin{abstract}
Roll-to-roll manufacturing requires precise tension and velocity control to ensure product quality, yet controller commissioning and adaptation remain time-intensive processes dependent on expert knowledge. This paper presents an LLM-assisted multi-agent framework that automates control system design and adaptation for R2R systems while maintaining safety. The framework operates through five phases: system identification from operational data, automated controller selection and tuning, sim-to-real adaptation with safety verification, continuous monitoring with diagnostic capabilities, and periodic model refinement.  Validation with a lab-scale R2R system demonstrates successful tension regulation and velocity tracking under significant model uncertainty, with the framework achieving performance convergence through iterative adaptation. Specifically, the framework achieves 55.7\% and 82.4\% RMSE reductions in tension control and velocity tracking respectively, outperforming an MPC baseline in both scenarios. The approach reduces manual tuning effort while providing transparent diagnostic information for maintenance planning, offering a practical pathway for integrating AI-assisted automation in manufacturing control systems.
\end{abstract}


\cortext[cor1]{Corresponding author. Tel.: +1-512-471-7174}

\end{frontmatter}

\begin{framed}
\noindent\textbf{Nomenclature}\par\vspace{0.5\baselineskip}
\begin{deflist}[$P_{\text{baseline}}$]
\defitem{$E$}\defterm{Modulus (N/m$^2$)}
\defitem{$A$}\defterm{Cross-sectional area (m$^2$)}
\defitem{$R_i$}\defterm{Radius of roller $i$ (m)}
\defitem{$J_i$}\defterm{Moment of inertia of roller $i$ (kg$\cdot$m$^2$)}
\defitem{$f_i$}\defterm{Friction coefficient of motor $i$ (N$\cdot$m$\cdot$s$\cdot$rad$^{-1}$)}
\defitem{$L_i$}\defterm{Length of web span $i$ (m)}
\defitem{$b_i$}\defterm{Disturbance coefficient of motor $i$}
\defitem{$\mathbf{x}(t)$}\defterm{State vector}
\defitem{$\mathbf{u}(t), \mathbf{u}_k$}\defterm{Control input vector}
\defitem{$\mathbf{y}(t), \mathbf{y}_k$}\defterm{Measured output vector}
\defitem{$T_i$}\defterm{Web tension in span $i$ (N)}
\defitem{$v_i$}\defterm{Linear velocity of roller $i$ (m$\cdot$s$^{-1}$)}
\defitem{$v_0$}\defterm{Unwinding velocity (m$\cdot$s$^{-1}$)}
\defitem{$v_i^r$}\defterm{Reference velocity of roller $i$ (m$\cdot$s$^{-1}$)}
\defitem{$\omega_i$}\defterm{Angular velocity of roller $i$ (rad$\cdot$s$^{-1}$)}
\defitem{$u_i$}\defterm{Control torque applied to motor $i$ (N$\cdot$m)}
\defitem{$\boldsymbol{\theta}$}\defterm{System parameter vector}
\end{deflist}
\end{framed}


\section{Introduction}

Advanced Roll-to-roll (R2R) manufacturing enables high-volume, cost-effective production of flexible electronics, printed sensors, and functional films \cite{greener2018roll}. Maintaining precise web tension and velocity control remains critical for product quality \citep{yan2020web,li2025adaptive}, with tension variations directly causing defects such as wrinkles, misalignment, and material damage \citep{brandenburg1976new}. The core manufacturing challenge lies in managing strongly coupled tension-velocity dynamics while accommodating time-varying parameters from changing roll radius, material property variations, and environmental conditions \citep{saad2000multivariable}. Current industrial practice relies on experienced operators and control engineers for manual tuning, requiring time and expertise \cite{niu2022process}—particularly problematic when transitioning between product grades or commissioning new production lines.

Large language models (LLMs) offer promising capabilities for automating engineering tasks through natural language reasoning and code generation \citep{ wang2024agents}. Retrieval-Augmented Generation (RAG) further enhances these capabilities by enabling LLMs to dynamically access external knowledge bases \citep{lewis2020rag}. Unlike conventional LLMs that rely only on static pre-training, RAG incorporates dynamic retrieval during inference \citep{gao2023retrieval}, improving contextual accuracy and reducing wrong outputs by grounding responses in domain-specific information. However, their application to manufacturing control faces critical practical barriers: ensuring operational safety and constraint satisfaction \citep{kirchner2025generating}, addressing the gap between simulation models and physical systems \citep{samak2025digital}, and maintaining transparent decision-making required in production environments \citep{brintrup2023trustworthy, baptista2025large}. The question for manufacturing practitioners is whether LLM capabilities can be harnessed to reduce commissioning time and operational burden while maintaining the reliability standards required in industrial settings.

This work proposes an LLM-assisted framework for R2R control that approaches these manufacturing requirements through simulation-validated adaptation. The framework employs specialized agents for system identification, controller design, adaptive tuning, and process monitoring—each operating within rigorous safety constraints. All proposed control modifications undergo simulation validation before deployment to production equipment. The approach delivers: (1) RAG-based knowledge infrastructure that grounds LLM reasoning in control theory and R2R manufacturing domain expertise; (2) automated controller selection and tuning methodology with built-in safety verification; (3) intelligent monitoring with implementation guidelines for manufacturing practitioners; (4) demonstration of successful deployment despite significant model uncertainty.

\section{Related Work}

\subsection{Traditional and AI-Driven Control in R2R Systems}
Classical R2R control relies on PID, LQR and MPC variants to regulate web tension and transport speed under strong coupling and changing roll radii. Foundational modeling and robust control for winding systems established today’s tension/velocity decoupling strategies \citep{koc2002winding}, while industrial studies refined adaptive PI/PID implementations to cope with operating-point drift \citep{raul2015adaptivePI,chen2017fuzzy}. Recent work emphasizes advanced control methods for R2R systems \citep{martin2024stabilization,martin2025sequential} and physics-consistent tension models validated on pilot lines \citep{jeong2021tension,he2024multispan}. The central role of monitoring and feedback design for achieving product quality and system robustness has been presented \citep{martin2022hinfty,martin2025hinfty}. On the data-driven side, recent R2R studies have demonstrated robust constrained tension 
regulation \citep{chen2023robustR2R} and AI-assisted tension field reconstruction 
\citep{gafurov2024webtension,gafurov2025aidt}. A recent review of web-tension control 
further identifies hybrid controllers as a promising direction \citep{martin2025review}.

\subsection{LLM for Control Design}
LLMs increasingly act as planners, code synthesizers, and supervisors for control systems. Agentic frameworks automate various aspects of control design, including requirement parsing, loop-shaping, and gain search, then call numerical solvers for verification \citep{guo2024controlagent,zahedifar2025llmac}. In cyber-physical systems, LLM agents co-design objective-oriented controllers and evolve control structures \citep{cui2024powercontrol,cui2025gencontrol}. For safety-critical applications, LLMs translate natural-language specifications into formal artifacts or symbolic-control pipelines \citep{bayat2025symbolic}. Recent work extends LLM-based control to robust synthesis via linear matrix inequalities (LMI) \citep{li2025natural} and adaptive model order reduction for control design \citep{li2025aurora}. These approaches show promise in automating control design tasks. However, practical deployment in industrial manufacturing remains limited, motivating frameworks that address safety validation.

\subsection{LLM in Manufacturing Applications}
LLMs in manufacturing-focused frameworks have enabled data integration, decision support, and workflow automation across processes and shop-floor information technology \citep{garcia2024framework,li2024llm4mfg}. LLM agents have been combined with digital twin resources to aid planning, diagnosis, and human-machine collaboration, while focusing primarily on analytics and high-level decision-making \citep{xia2023flexible,chen2025llmdt,ouerghemmi2025enhanceddt}. LLM-enabled knowledge extraction has been reported for process engineering \citep{liu2025ner}, Industrial Internet of Things (IIoT) \citep{gautam2025iiot}, and agentic fault-handling workflows \citep{gill2025faultllmdt}. However, most existing manufacturing applications focus on supervisory control, such as task sequencing, resource allocation, and fault detection, rather than control system design itself. This gap remains largely because controller design requires careful mathematical reasoning about system dynamics and stability, depends on closed-loop setups for iterative tuning using simulations or process data, and does not naturally fit the language-based workflows that generate training data for LLMs. As a result, prior work tends to address what control actions to take—such as selecting setpoints—without tackling how to design controllers that can reach those setpoints with the desired dynamic behavior. These limitations motivate an LLM-assisted framework for R2R manufacturing that unifies controller design, parameter tuning, and sim-to-real adaptation.

\section{Proposed Framework}
\label{sec:methodology}

Integrating LLMs into manufacturing control represents significant challenges, including ensuring safety despite probabilistic reasoning, maintaining transparent decisions for regulatory compliance, and bridging general-purpose models with domain-specific control expertise. We propose a framework to address these through multi-agent collaboration, simulation-based safety validation, and retrieval-augmented domain knowledge integration.

\begin{figure}[htbp]
    \centering
    \includegraphics[width=0.4\textwidth]{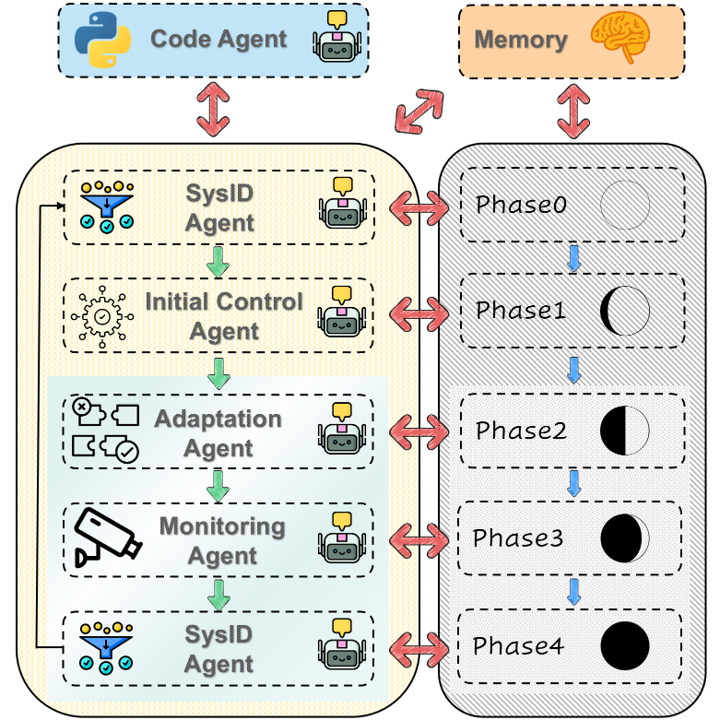}
    \caption{LLM-assisted R2R control framework. All LLM-proposed modifications must pass simulation validation before deployment.}
    \label{fig:framework_overview}
\end{figure}

The framework operates through five phases: system initialization (Phase 0), offline controller design (Phase 1), real system deployment (Phase 2), stable operation with monitoring (Phase 3), and periodic model updates (Phase 4). Unlike conventional automation, this implements constrained autonomy—LLMs generate strategies, simulation validates safety, and humans retain oversight. Figure \ref{fig:framework_overview} shows the closed-loop architecture where operational data informs simulation refinement, while LLMs orchestrate design, tuning, and adaptation.

\subsection{Multi-Agent Architecture and RAG Integration}

The framework's intelligence emerges from specialized agent collaboration rather than monolithic reasoning. This design addresses a key limitation of applying general-purpose LLMs to control: the need for both broad reasoning capabilities and deep domain expertise at different decision points. 

The framework employs five specialized agents with role-specific prompts and domain knowledge access: SysID Agent (system identification), Initial Control Agent (controller tuning and selection), Adaptation Agent (sim-to-real tuning), Monitoring Agent (monitoring and diagnostics), and Code Agent (algorithm application and validated execution). Each interaction is logged with requesting agent, task specification, retrieved knowledge, and validation results. 

\subsubsection{Prompt Engineering for Control Domain}

Control system prompts differ from general LLM applications in four ways: (1) state physical limits clearly to reduce hallucinations, (2) organize reasoning steps from data analysis to testing, (3)  mandate validation before deployment, and (4) define protocols for human intervention. As detailed in \ref{sec:agent_prompts}.

\subsubsection{RAG for Domain Knowledge Grounding}

A major distinction between human engineers and LLMs is knowledge access: humans leverage years of accumulated experience, while LLMs carry the risk of generating control strategies that appear reasonable yet are dangerous. RAG \citep{gao2023retrieval,gautam2025iiot} bridges this gap by dynamically grounding LLM reasoning in verified domain knowledge.

\begin{figure*}[htbp]
    \centering
    \includegraphics[width=0.75\textwidth]{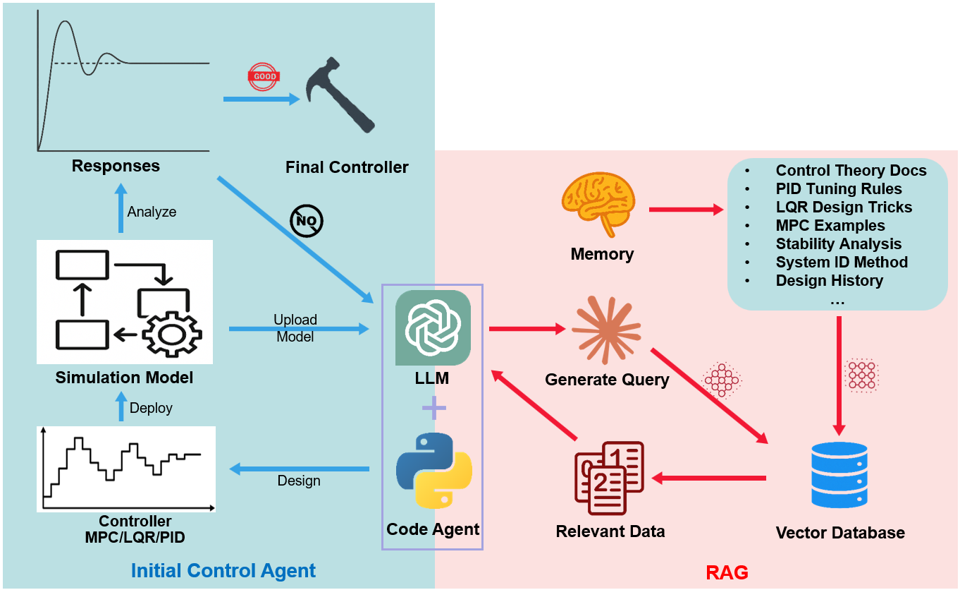}
    \caption{Initial Control Agent with RAG integration. All agents access this infrastructure; detailed here for clarity.}
    \label{fig:InitialControl}
\end{figure*}

The knowledge base provides three hierarchical tiers: (1) control theory fundamentals (e.g., stability criteria and design methodologies), (2) R2R best practices (e.g., tension control strategies and commissioning procedures), and (3) system-specific documentation (e.g., equipment specs and operational history). When agents query, multi-stage retrieval performs semantic embedding, hierarchical searching, relevance ranking, and conflict resolution.This enables recall from thousands of documents and consistency verification across sources. Figure \ref{fig:InitialControl} illustrates RAG integration for the Initial Control Agent.

\subsection{Phase 0: System Identification (SysID Agent)}

With the architectural foundation established, we now detail how each phase leverages LLM capabilities. Phase 0 demonstrates how LLMs automate the traditionally expert-driven task of model construction.

\begin{figure}[htbp]
    \centering
    \includegraphics[width=0.4\textwidth]{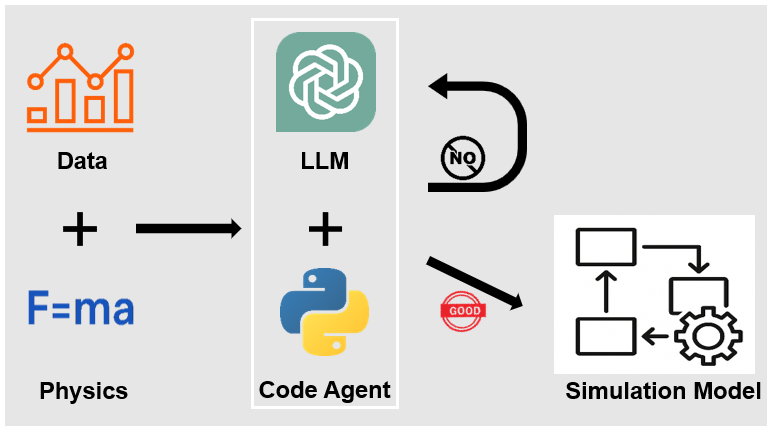}
    \caption{SysID Agent workflow}
    \label{fig:sysID}
\end{figure}

This phase constructs a  simulation model from operational data and physical principles, serving dual purposes: enabling offline controller design and providing safety filter validation. Figure \ref{fig:sysID} shows the workflow of SysID Agent.

Historical data $\mathcal{D} = \{(u_k, y_k, t_k)\}_{k=1}^{N}$ is collected, where $u_k \in \mathbb{R}^{m}$ represents control inputs, $y_k \in \mathbb{R}^{p}$ measured outputs, and $t_k$ timestamps. The system dynamics follow a nonlinear state-space representation:
\begin{equation}
\dot{\mathbf{x}}(t) = f(\mathbf{x}(t), \mathbf{u}(t), \boldsymbol{\theta}), \quad \mathbf{y}(t) = h(\mathbf{x}(t), \boldsymbol{\theta})
\label{eq:system_dynamics}
\end{equation}
where $x(t) \in \mathbb{R}^{n}$ contains web tensions and roller velocities, $\theta$ represents physical parameters, and functions $f(\cdot)$ and $h(\cdot)$ represent the physics of web transport.

The SysID Agent combines data-driven estimation with physics-informed structure. Upon receiving data, it queries RAG for R2R identification practices and analyzes data characteristics to formulate a strategy. The agent automatically translates physical principles into optimization constraints. Rather than relying on a single identification method, the agent implements multiple approaches including recursive least squares, prediction error methods, and subspace identification.

The identification problem minimizes prediction error while regularizing toward prior knowledge from equipment specifications:
\begin{equation}
\boldsymbol{\theta}^* = \arg\min_{\boldsymbol{\theta}} \sum_{k=1}^{N} \|\mathbf{y}_k - \hat{\mathbf{y}}_k(\boldsymbol{\theta})\|_2^2 + \lambda \|\boldsymbol{\theta} - \boldsymbol{\theta}_{\text{prior}}\|_2^2
\label{eq:identification_objective}
\end{equation}
The regularization term prevents overfitting and constrains parameters to physically meaningful ranges.

The control objective is defined as:
\begin{equation}
\min_{\mathbf{u}(t)} \quad J = \int_{0}^{T_f} \left[(\mathbf{y}(t) - \mathbf{y}_{\text{ref}}(t))^\top \mathbf{Q} (\mathbf{y}(t) - \mathbf{y}_{\text{ref}}(t)) + \mathbf{u}(t)^\top \mathbf{R} \mathbf{u}(t)\right] dt
\label{eq:control_objective}
\end{equation}
subject to physical and operational constraints:
\begin{equation}
\mathbf{x}_{\min} \le \mathbf{x}(t) \le \mathbf{x}_{\max}, \quad \mathbf{u}_{\min} \le \mathbf{u}(t) \le \mathbf{u}_{\max}
\label{eq:constraints}
\end{equation}
where $\mathbf{Q}$ and $\mathbf{R}$ are weighting matrices, and $\mathbf{y}_{\text{ref}}(t)$ is the reference trajectory.

\subsection{Phase 1: Controller Design and Selection (Initial Control Agent)}

Building on the validated model, this phase evaluates three architectures. The LLM first analyzes system characteristics such as tension-velocity coupling strength and dynamic time scales to match appropriate controller architectures. This includes determining whether to linearize the nonlinear dynamics around operating points for controllers like LQR, or to use the full nonlinear model for nonlinear control approaches. It then reasons about parameter spaces using control theory principles to guide the tuning process. Finally, it evaluates trade-offs between competing objectives—tracking accuracy, settling speed, energy consumption, and robustness.

The PID controller is implemented as:
\begin{equation}
\mathbf{u}(t) = \mathbf{K}_p \mathbf{e}(t) + \mathbf{K}_i \int_0^t \mathbf{e}(\tau) d\tau + \mathbf{K}_d \frac{d\mathbf{e}(t)}{dt}
\label{eq:pid}
\end{equation}
where $\mathbf{e}(t) = \mathbf{y}_{\text{ref}}(t) - \mathbf{y}(t)$ is the tracking error, and $\mathbf{K}_p$, $\mathbf{K}_i$, $\mathbf{K}_d$ are the proportional, integral, and derivative gain matrices.

The MPC optimization problem at time step $k$ is formulated as:
\begin{equation}
\min_{\mathbf{u}} \sum_{i=0}^{N_p-1} \left[\|\mathbf{y}(k+i|k) - \mathbf{y}_{\text{ref}}(k+i)\|_{\mathbf{Q}}^2 + \|\mathbf{u}(k+i)\|_{\mathbf{R}}^2\right]
\label{eq:mpc}
\end{equation}
subject to system dynamics and constraints over prediction horizon $N_p$.

For the linearized system $\dot{\mathbf{x}} = \mathbf{A}\mathbf{x} + \mathbf{B}\mathbf{u}$, the LQR control law is:
\begin{equation}
\mathbf{u}(t) = -\mathbf{K}\mathbf{x}(t), \quad \mathbf{K} = \mathbf{R}^{-1}\mathbf{B}^\top\mathbf{P}
\label{eq:lqr}
\end{equation}
where $\mathbf{P}$ solves the continuous-time algebraic Riccati equation:
\begin{equation}
\mathbf{A}^\top\mathbf{P} + \mathbf{P}\mathbf{A} - \mathbf{P}\mathbf{B}\mathbf{R}^{-1}\mathbf{B}^\top\mathbf{P} + \mathbf{Q} = \mathbf{0}
\end{equation}

For each controller architecture, the LLM employs iterative reasoning to tune hyperparameters $\boldsymbol{\phi} = \{\mathbf{K}_p, \mathbf{K}_i, \mathbf{K}_d\}$ for PID, or equivalent parameters for MPC and LQR. The tuning process is described in Algorithm~\ref{alg:hyperparameter_tuning} below, where the LLM analyzes simulation responses and adjusts parameters to optimize the control objective defined in Equation~\eqref{eq:control_objective}.

\begin{algorithm}[htbp]
\caption{Initial Control Agent}
\label{alg:hyperparameter_tuning}
\begin{algorithmic}[1]
\REQUIRE Simulation model $\mathcal{S}_{\text{sim}}$, controller type $C$, objective $J$
\ENSURE Optimized hyperparameters $\boldsymbol{\phi}^*$
\STATE LLM retrieves tuning guidelines from RAG, initializes $\boldsymbol{\phi}_0$
\STATE $i \leftarrow 0$, $\text{max\_iterations} \leftarrow 50$
\WHILE{$i < \text{max\_iterations}$ AND not converged}
    \STATE Code Agent: Execute simulation with $C(\boldsymbol{\phi}_i)$, compute $\mathbf{P}(t)$
    \STATE LLM: Diagnose limitations, query RAG, generate $\boldsymbol{\phi}_{i+1}$ with justification
    \STATE Validate stability and constraints
    \IF{$J(\boldsymbol{\phi}_{i+1}) < J(\boldsymbol{\phi}_i)$ AND validated}
        \STATE Accept $\boldsymbol{\phi}_{i+1}$, log reasoning
    \ELSE
        \STATE LLM revises strategy
    \ENDIF
    \STATE $i \leftarrow i + 1$
\ENDWHILE
\STATE \RETURN $\boldsymbol{\phi}^*$ with tuning history
\end{algorithmic}
\end{algorithm}

Performance metrics evaluated here include root mean square error quantifying tracking accuracy:
\begin{equation}
e_{\text{RMSE}} = \sqrt{\frac{1}{N}\sum_{k=1}^{N}\|\mathbf{y}_k - \mathbf{y}_{\text{ref},k}\|_2^2}
\label{eq:rmse}
\end{equation}
settling time measuring speed of convergence to steady-state:
\begin{equation}
t_s = \min\{t : \|\mathbf{y}(\tau) - \mathbf{y}_{\infty}\|_2 \leq 0.02\|\mathbf{y}_{\infty}\|_2, \; \forall \tau \geq t\}
\label{eq:settling_time}
\end{equation}
overshoot percentage indicating stability margin and oscillatory behavior:
\begin{equation}
OS = \frac{\|\max(\mathbf{y}(t)) - \mathbf{y}_{\infty}\|_2}{\|\mathbf{y}_{\infty}\|_2} \times 100\%
\label{eq:overshoot}
\end{equation}
control effort measuring energy consumption and actuator stress:
\begin{equation}
U_{\text{total}} = \int_{0}^{T_f} \mathbf{u}(t)^\top \mathbf{u}(t) \, dt
\label{eq:control_effort}
\end{equation}
and robustness quantifying worst-case performance under model uncertainty:
\begin{equation}
\mathcal{R} = \max_{\|\Delta\boldsymbol{\theta}\| \leq \epsilon} \|\mathbf{y}(t; \boldsymbol{\theta} + \Delta\boldsymbol{\theta}) - \mathbf{y}(t; \boldsymbol{\theta})\|_{\infty}
\label{eq:robustness}
\end{equation}

Define the performance vector as $\mathbf{P}(t) = [e_{\text{RMSE}}(t), OS(t), t_s(t)]^\top$, which characterizes controller performance across accuracy, speed, stability and efficiency.

Selection via weighted score balancing multiple objectives:
\begin{equation}
S = \mathbf{w}^\top [e_{\text{RMSE}}, t_s, OS, U_{\text{total}}, \mathcal{R}, C_{\text{comp}}]^\top
\label{eq:performance_score}
\end{equation}
where weights $\mathbf{w} = [w_1, w_2, w_3, w_4, w_5, w_6]^\top$ reflect manufacturing priorities and $C_{\text{comp}}$ accounts for computational feasibility.

\begin{equation}
C^* = \arg\min_{C \in \{\text{PID, MPC, LQR}\}} S(C)
\end{equation}

This optimal controller $C^*$ with its tuned parameters $\boldsymbol{\phi}^*$ is then prepared for deployment to the real system in Phase 2.

\subsection{Phase 2: Sim-to-Real Adaptation (Adaptation Agent)}

\begin{figure*}[htbp]
    \centering
    \includegraphics[width=0.75\textwidth]{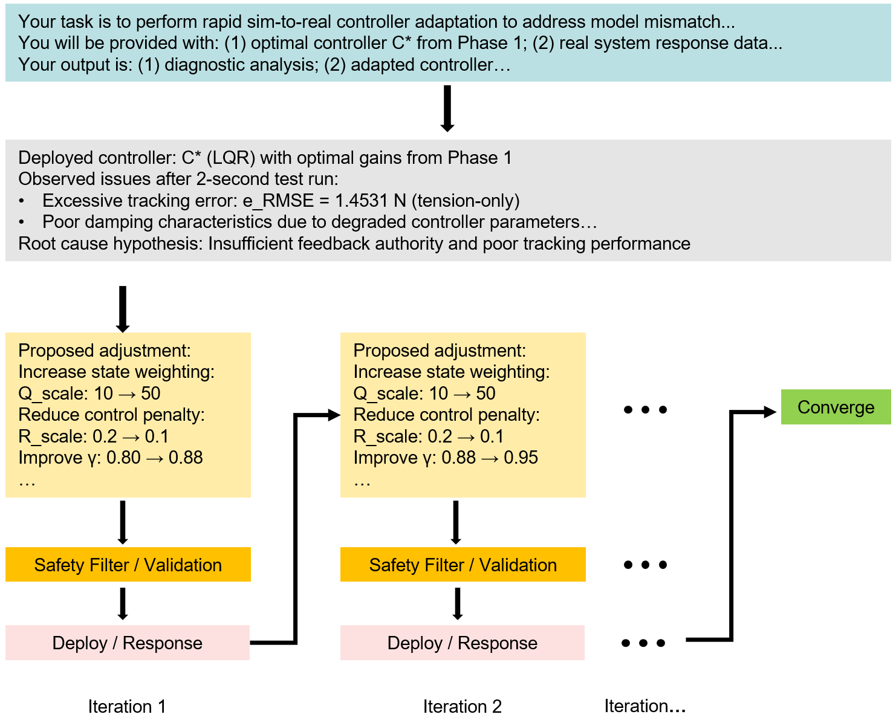}
    \caption{Adaptation Agent with mandatory safety filter. Simulation-in-the-loop ensures zero unsafe deployments.}
    \label{fig:Adapt}
\end{figure*}

Phase 2 addresses sim-to-real gap through iterative adaptation, which is constrained autonomy: LLM proposes modifications, but every change must pass safety validation. Figure \ref{fig:Adapt} shows the workflow.

\subsubsection{Safety Filter}

A fundamental challenge is that LLMs generate probabilistic suggestions without safety guaranties. Human engineers rely on physical intuition to avoid dangerous choices; LLMs lack this. Our safety filter implements mandatory pre-validation: every proposed modification must demonstrate safe operation in simulation before hardware deployment.

The filter evaluates three criteria. Constraint satisfaction verifies that all operational limits are maintained throughout the control horizon:
\begin{equation}
\mathcal{C}_{\text{safe}} = \begin{cases}
\mathbf{u}_{\min} \leq \mathbf{u}_i(t) \leq \mathbf{u}_{\max}, & \forall i, t \\
|\dot{\mathbf{u}}_i(t)| \leq \dot{\mathbf{u}}_{\max}, & \forall i, t
\end{cases}
\label{eq:constraint_verification}
\end{equation}
where the rate constraint on control derivatives prevents actuator damage from abrupt torque changes that could occur with aggressive LLM-suggested parameters.

Performance improvement ensures that modifications enhance rather than degrade tracking accuracy, settling behavior, and overshoot:
\begin{equation}
\mathcal{P}_{\text{improved}} = \|\mathbf{P}_{\text{proposed}}(t)\|_2 < \|\mathbf{P}_{\text{current}}(t)\|_2
\label{eq:performance_verification}
\end{equation}
This metric prevents the LLM from suggesting changes that satisfy constraints but worsen control quality—a scenario human operators would recognize, but requires explicit verification for LLMs.

Stability margins assess robustness through worst-case performance under parametric uncertainty:
\begin{equation}
\mathcal{S}_{\text{robust}} = \max_{\|\Delta\boldsymbol{\theta}\| \leq \epsilon} \|\mathbf{y}(t; \boldsymbol{\theta} + \Delta\boldsymbol{\theta}) - \mathbf{y}_{\text{ref}}(t)\|_{\infty} < \delta_{\text{tol}}
\label{eq:stability_verification}
\end{equation}
Robustness is important because simulation models inevitably contain uncertainty; controllers must maintain stability despite model mismatch.

Deployment approval requires all three criteria to be satisfied simultaneously:
\begin{equation}
\text{Approve} = \mathcal{C}_{\text{safe}} \land \mathcal{P}_{\text{improved}} \land \mathcal{S}_{\text{robust}}
\label{eq:safety_approval}
\end{equation}
providing a safety barrier between LLM suggestions and physical implementation.

Safety filter provides capabilities difficult for humans: testing across comprehensive scenarios in seconds, zero fatigue across iterations, quantitative stability margins versus qualitative assessment, and automatic rollback preparation.

\subsubsection{Adaptation Process}

The adaptation process (Algorithm~\ref{alg:sim_to_real}) implements structured iteration that begins with deploying the Phase 1 controller $C^*(\boldsymbol{\phi}^*)$ to the real system. The Adaptation Agent then analyzes the performance gap between simulation and reality to diagnose the root cause of mismatch. Based on this diagnosis, it proposes parameter adjustments that must pass the simulation safety filter before deployment to hardware. This cycle repeats until the controller meets the convergence criteria defined in Equation~\eqref{eq:convergence_criteria}.

\begin{algorithm}[htbp]
\caption{Adaptation Agent}
\label{alg:sim_to_real}
\begin{algorithmic}[1]
\REQUIRE Controller $C^*(\boldsymbol{\phi}^*)$, real system $\mathcal{S}_{\text{real}}$, thresholds, simulation $\mathcal{S}_{\text{sim}}$
\ENSURE Adapted controller $C_{\text{adapted}}$
\STATE Deploy $C^*(\boldsymbol{\phi}^*)$ to $\mathcal{S}_{\text{real}}$
\STATE $\text{iteration} \leftarrow 0$
\REPEAT
    \STATE Monitor tension $T(t)$ and velocity $v(t)$, compute $\mathbf{P}_{\text{real}}(t)$
    \IF{$\mathbf{P}_{\text{real}}(t)$ satisfies Eq.~\eqref{eq:convergence_criteria}}
        \STATE Performance met, exit
    \ENDIF
    \STATE LLM: Analyze gap $\Delta \mathbf{P}$, query RAG for causes, diagnose root cause
    \STATE LLM: Generate $\Delta\boldsymbol{\phi}$ with justification
    \STATE {Safety Filter:} Test $C(\boldsymbol{\phi} + \Delta\boldsymbol{\phi})$ in $\mathcal{S}_{\text{sim}}$
    \STATE Verify per Eqs.~\eqref{eq:constraint_verification}-\eqref{eq:stability_verification}
    \IF{approved per Eq.~\eqref{eq:safety_approval}}
        \STATE Deploy $\boldsymbol{\phi} \leftarrow \boldsymbol{\phi} + \Delta\boldsymbol{\phi}$ to $\mathcal{S}_{\text{real}}$
        \STATE Log: parameters, reasoning, results
    \ELSE
        \STATE Reject, LLM revises
    \ENDIF
    \STATE $\text{iteration} \leftarrow \text{iteration} + 1$
\UNTIL{converged OR max iterations}
\STATE \RETURN $C_{\text{adapted}}$ with history
\end{algorithmic}
\end{algorithm}

Convergence criteria define when real system performance meets production requirements:
\begin{equation}
\mathbf{P}_{\text{real}}(t) < \mathbf{P}_{\text{threshold}} = [0.05 \cdot T_{\text{ref}}, 20\%, 2.0 \text{s}]^\top.
\label{eq:convergence_criteria}
\end{equation}

\subsection{Phase 3: Intelligent Monitoring (Monitoring Agent)}

\begin{figure}[htbp]
    \centering
    \includegraphics[width=0.4\textwidth]{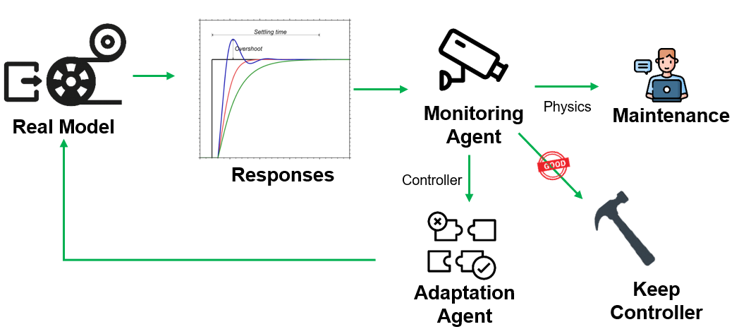}
    \caption{Monitoring Agent dual-layer architecture}
    \label{fig:Monitor}
\end{figure}

Following successful adaptation, Phase 3 implements continuous monitoring with autonomous diagnostics. LLMs have capabilities beyond human monitoring, including (1) 24/7 vigilance without fatigue, (2) multi-hypothesis reasoning with confidence scores avoiding cognitive bias, and (3) cross-domain knowledge integration from RAG. Unlike threshold-based alarms, the Intelligent Monitoring agent performs dual-layer analysis: detecting degradation and diagnosing root causes—distinguishing control-adjustable issues from physical maintenance needs. Figure \ref{fig:Monitor} shows the architecture.

Layer 1 detects degradation by computing the Euclidean distance from baseline performance:
\begin{equation}
\Delta \mathbf{P}(t) = \|\mathbf{P}(t) - \mathbf{P}_{\text{baseline}}\|_2 > 2\sigma_{\text{baseline}}
\label{eq:degradation_threshold}
\end{equation}
using a statistical threshold that adapts to normal process variability captured in $\sigma_{\text{baseline}}$.

Layer 2 diagnoses causes, such as material property variations, mechanical degradation (e.g., bearing wear and belt slippage), sensor issues (e.g., drift and noise), environmental factors, and process disturbances. The agent formulates hypotheses by matching observed patterns against RAG-retrieved failure modes, assigning confidence based on evidence strength:
\begin{equation}
H_j = \{\text{hypothesis}_j, \text{confidence}_j, \text{evidence}_j, \text{action}_j\}
\end{equation}

\begin{algorithm}[htbp]
\caption{Monitoring Agent}
\label{alg:adaptive_adjustment}
\begin{algorithmic}[1]
\REQUIRE Real-time data stream $\mathcal{D}_{\text{live}}$, baseline performance $\mathbf{P}_{\text{baseline}}$
\ENSURE Adapted controller or maintenance alert
\STATE Monitor $\mathbf{P}(t)$ continuously
\IF{degradation detected per Eq.~\eqref{eq:degradation_threshold}}
    \STATE Route to Adaptation Agent for control adaptation
    \IF{adaptation successful: performance recovers to $\|\Delta \mathbf{P}\|_2 < \sigma_{\text{baseline}}$}
        \STATE Deploy adapted controller
        \STATE Perform Layer 2 diagnostics to identify root cause
        \STATE Log findings for predictive maintenance
    \ELSE
        \STATE Send immediate maintenance alert 
        \STATE Perform Layer 2 diagnostics for root cause analysis
        \STATE Provide diagnostic evidence to maintenance team
        \STATE Continue with current controller pending repair
    \ENDIF
\ENDIF
\end{algorithmic}
\end{algorithm}

\subsection{Phase 4: Continuous Model Refinement (SysID Agent)}
During scheduled R2R system downtime or rest periods, accumulated operational data $\mathcal{D}_{\text{ops}} = \{(\mathbf{u}_k, \mathbf{y}_k, t_k)\}_{k=1}^{M}$ from Phases 2 and 3 is used to refine the simulation model. The simulation model parameters are re-identified following the same system identification and physics-informed construction procedures established in Phase 0, ensuring the simulation environment remains representative of the evolving real system dynamics and safety validation remains representative. Simultaneously, the controller parameters $\boldsymbol{\phi}_{\text{stable}}$ that achieved stable operation in Phase 3 are archived as the validated baseline configuration. For subsequent production runs, these archived controller parameters can be deployed directly to the real system.

\section{Validation Studies}
\label{sec:experiments}

This section validates the proposed LLM-assisted control framework through a simulation study of a R2R web handling system.

\subsection{System Description and Dynamic Model Formulation}

\begin{figure}[htbp]
    \centering
    \includegraphics[width=0.4\textwidth]{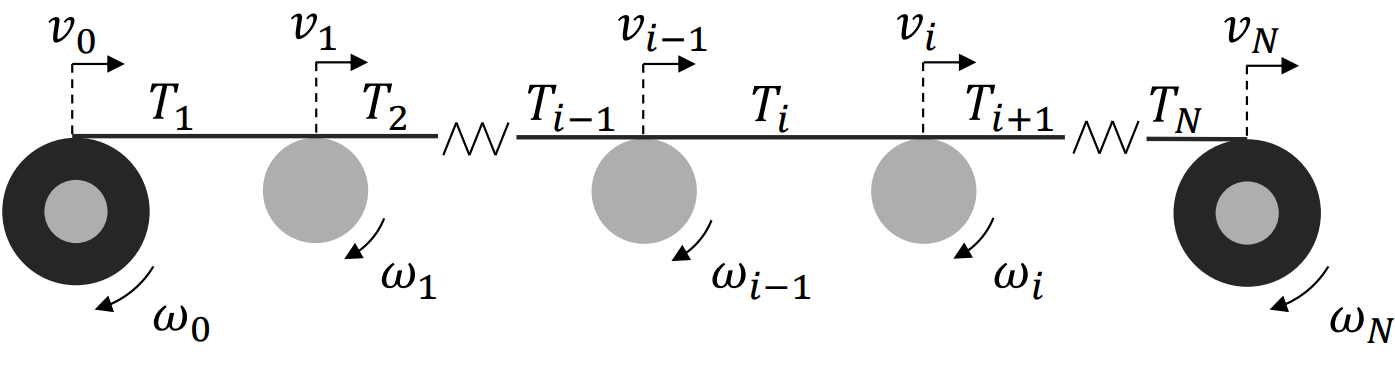}
    \caption{Schematic of a simplified R2R line}
    \label{fig:r2r_schematic}
\end{figure}

The first validation employs a multi-span R2R web handling system representative of industrial processes such as printing and coating. The system schematic and nomenclature are shown in Figure~\ref{fig:r2r_schematic}. Following established approaches \cite{Shelton1986, Martin2025a}, the model adopts three key assumptions:(1) Passive rollers are neglected, with only actuated motorized rollers considered \cite{Martin2025a}.
 (2) Web tension $T_i$ remains uniform within each span $i$ between adjacent rollers.
 (3) The no-slip condition $v_i = \omega_i R_i$ holds due to sufficiently high friction.

The system dynamics couple web tension evolution and roller velocity through differential equations. Web tension dynamics in span $i$ follow the viscoelastic relation \cite{Shelton1986}:
\begin{equation}
\frac{dT_i}{dt} = \frac{AE}{L_i}(v_i - v_{i-1}) + \frac{1}{L_i}(T_{i-1}v_{i-1} - T_i v_i)
\label{eq:web_tension_dynamics}
\end{equation}
where $A$ is web cross-sectional area, $E$ is elastic modulus, $L_i$ is span length, and $v_i$, $v_{i-1}$ are roller velocities. Roller velocity dynamics from torque balance are:
\begin{equation}
\frac{dv_i}{dt} = \frac{R_i^2}{J_i}(T_{i+1} - T_i) - \frac{f_i}{J_i}v_i + \frac{R_i}{J_i}u_i + b_i\xi_i(t)
\label{eq:roller_velocity_dynamics}
\end{equation}
where $R_i$ is roller radius, $J_i$ is moment of inertia, $f_i$ is viscous friction, $u_i$ is control torque, and $b_i\xi_i(t)$ captures stochastic disturbances. Parameters $\boldsymbol{\theta} = \{E, A, L_i, R_i, J_i, f_i\}$ are identified following Phase 0 procedures. The validation uses three parameter sets: real system(simulation) $\mathcal{S}_{\text{real}}$ generating operational data, identified model $\mathcal{S}_{\text{real,ID}}$ from Phase 0 system identification with 1.4-8.6\% estimation errors, and intentionally mismatched simulation model $\mathcal{S}_{\text{sim}}$ with 8.3-50\% deviations to test adaptation robustness under significant sim-to-real gap (Table~\ref{tab:system_parameters}). This conservative setting ensures generalizability—by testing under intentionally large model mismatches, successful adaptation here suggests the framework should perform well in typical real-world scenarios where gaps are smaller.

\begin{table}[htbp]
\centering
\caption{System Parameters: Real, Identified, and Simulation Models}
\label{tab:system_parameters}
\begin{tabular}{lccc}
\hline
\textbf{Symbol (Unit)} & \textbf{$\mathcal{S}_{\text{real}}$} & \textbf{$\mathcal{S}_{\text{real,ID}}$} & \textbf{$\mathcal{S}_{\text{sim}}$} \\
\hline
$EA$ (N) & 2200 & 2169.85 & 2400 \\
$J$ (kg$\cdot$m$^2$) & 1.0 & 1.078 & 0.95 \\
$R$ (m) & 0.035 & 0.038 & 0.04 \\
$f_i$ (N$\cdot$m$\cdot$s/rad) & 15.0 & 15.47 & 10.0 \\
$L$ (m) & 1.2 & 1.24 & 1.0 \\
Disturbance (s$\cdot$m$^{-1}$$\cdot$kg$^{-1}$) & $5\cdot10^{-2}$ & - & $10^{-2}$ \\
Number of spans & 6 & 6 & 6 \\
\hline
\end{tabular}
\end{table}

\subsection{Tension Control}

\paragraph{System Configuration}
The tension control scenario maintains setpoints $T_1 = 28$ N, $T_2 = 36$ N, $T_4 = 40$ N, $T_5 = 24$ N, and $T_6 = 32$ N, with web 3 executing a step change from 20 N to 44 N at $t = 1$ s to simulate process disturbances. Unwinding velocity is $v_0 = 0.01$ m/s, with reference roller velocities computed as:
\begin{equation}
v_i^{r}(t) = \frac{EA - T_{i-1}(t)}{EA - T_i(t)} v_{i-1}^{r}(t), \quad v_0^{r} = v_0
\label{eq:reference_velocity}
\end{equation}
The low unwinding velocity was chosen to focus on validating the LLM-assisted adaptation framework under controlled conditions. At higher transport speeds typical of production, the tension-velocity coupling becomes stronger and disturbance rejection 
more demanding, which would further exercise the adaptation loop.

\begin{figure}[htbp]
    \centering
    \includegraphics[width=0.5\textwidth]{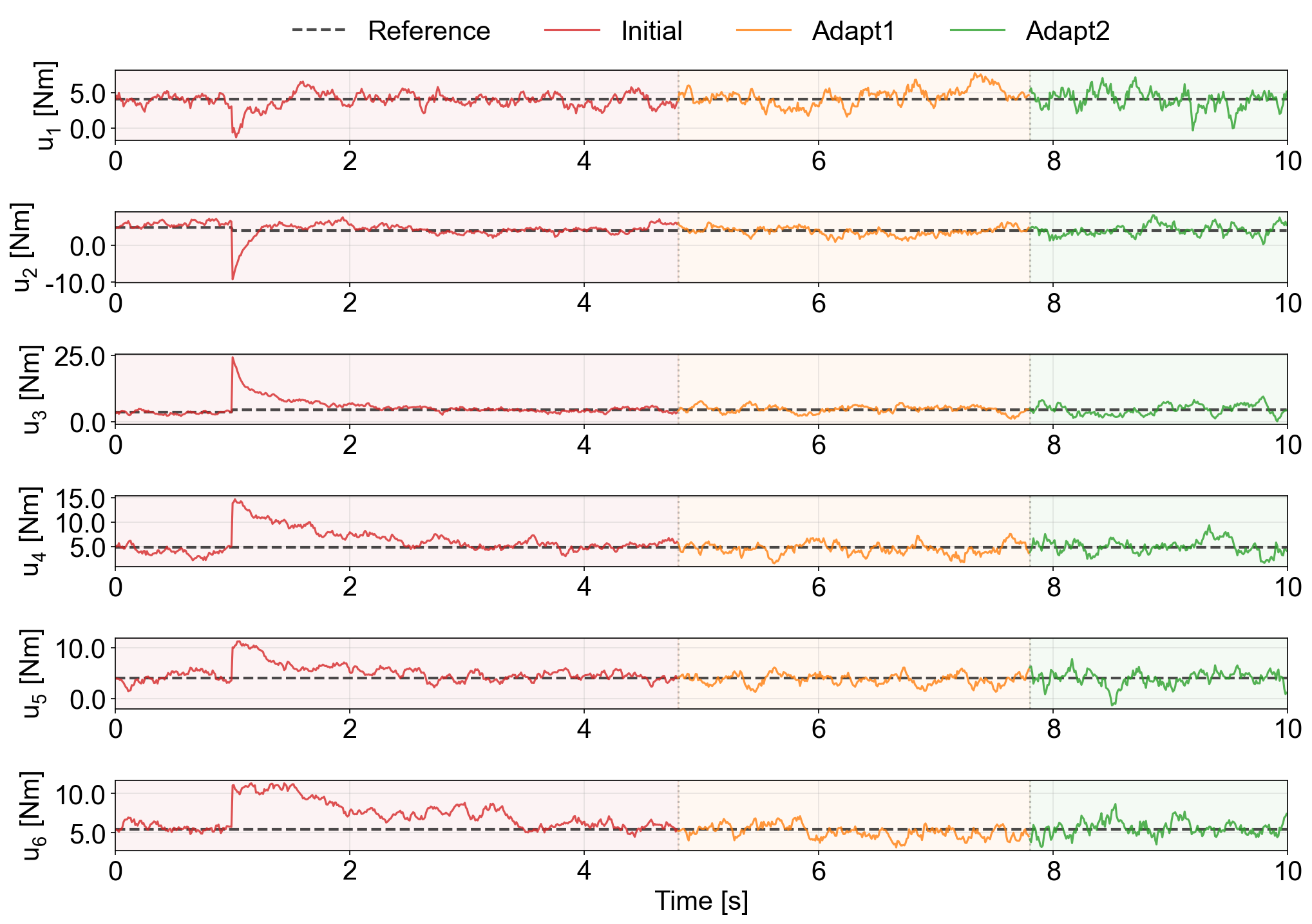}
    \caption{Control input trajectories during tension control}
    \label{fig:sequential_control}
\end{figure}

\begin{figure}[htbp]
    \centering
    \includegraphics[width=0.5\textwidth]{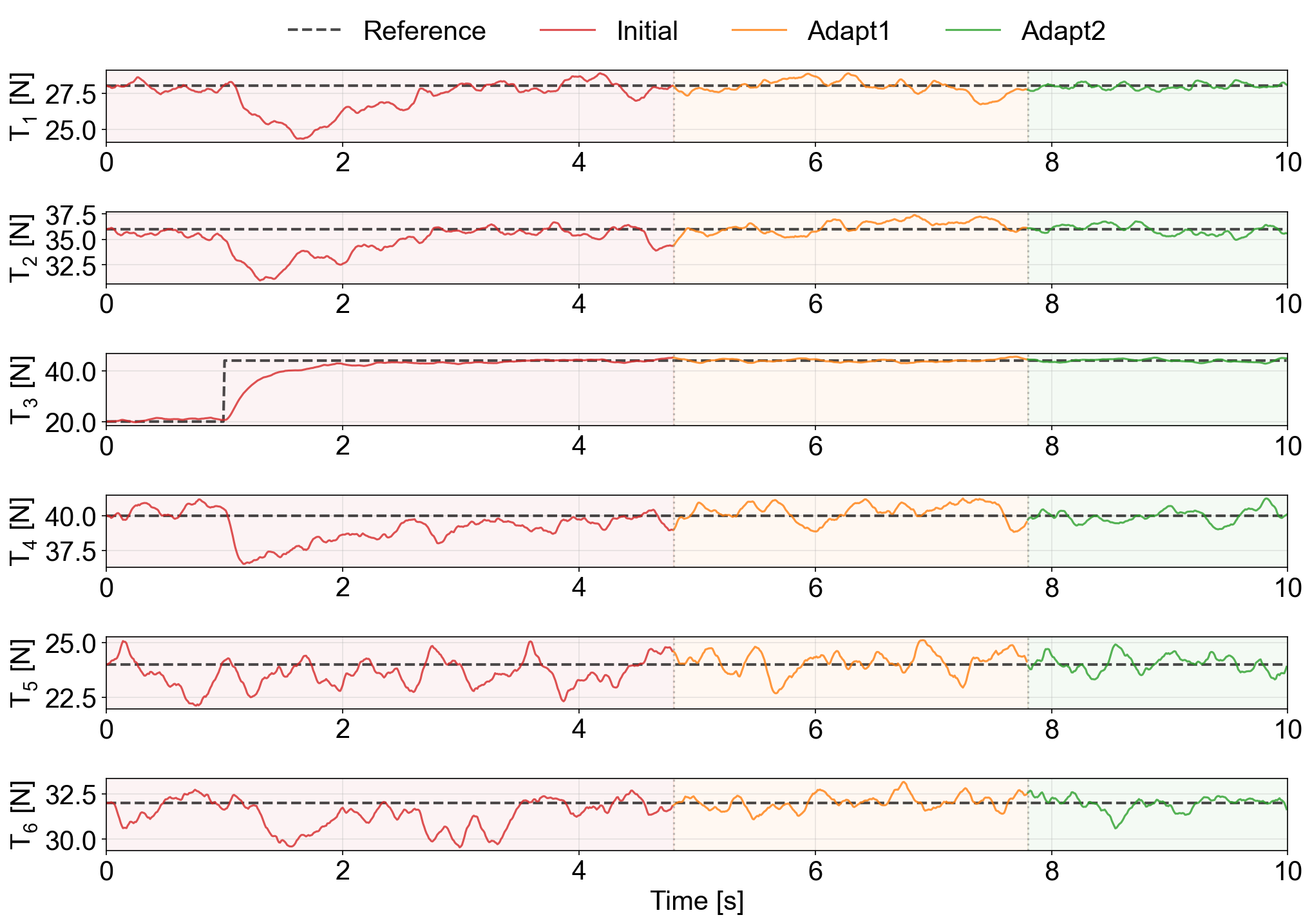}
    \caption{Tension trajectories during tension control}
    \label{fig:sequential_tensions}
\end{figure}

\paragraph{Controller Adaptation}
The Initial Control Design Agent selected LQR based on Table~\ref{tab:controller_comparison}. Figure~\ref{fig:sequential_control} and Figure~\ref{fig:sequential_tensions} show three sequential adaptation phases. Initial controller has large tracking errors and control oscillations upon initial deployment. While Adapt1,2 demonstrates improved stability with smooth control inputs.

\begin{table}[htbp]
\vspace{6pt}
\centering
\caption{Comparative Performance of Controller Architectures in Simulation Environment}
\label{tab:controller_comparison}
\begin{tabular}{lccccc}
\hline
\textbf{Type} & \textbf{$e_{\text{RMSE}}$ (N)} & \textbf{$t_s$ (s)} & \textbf{OS (\%)} & \textbf{$U_{\text{total}}$} & \textbf{$C_{\text{comp}}$ (ms)} \\
\hline
PID    & 0.4813 & 1.05 & 16.41 & 706.20 & 0.0082 \\
MPC    & 0.5843 & 1.18 & 13.89 & 703.64 & 0.2319 \\
LQR    & \textbf{0.3811} & 1.62 & 16.31 & 703.81 & \textbf{0.0002} \\
\hline
\end{tabular}
\end{table}

\begin{figure}[htbp]
    \centering
    \includegraphics[width=0.42\textwidth]{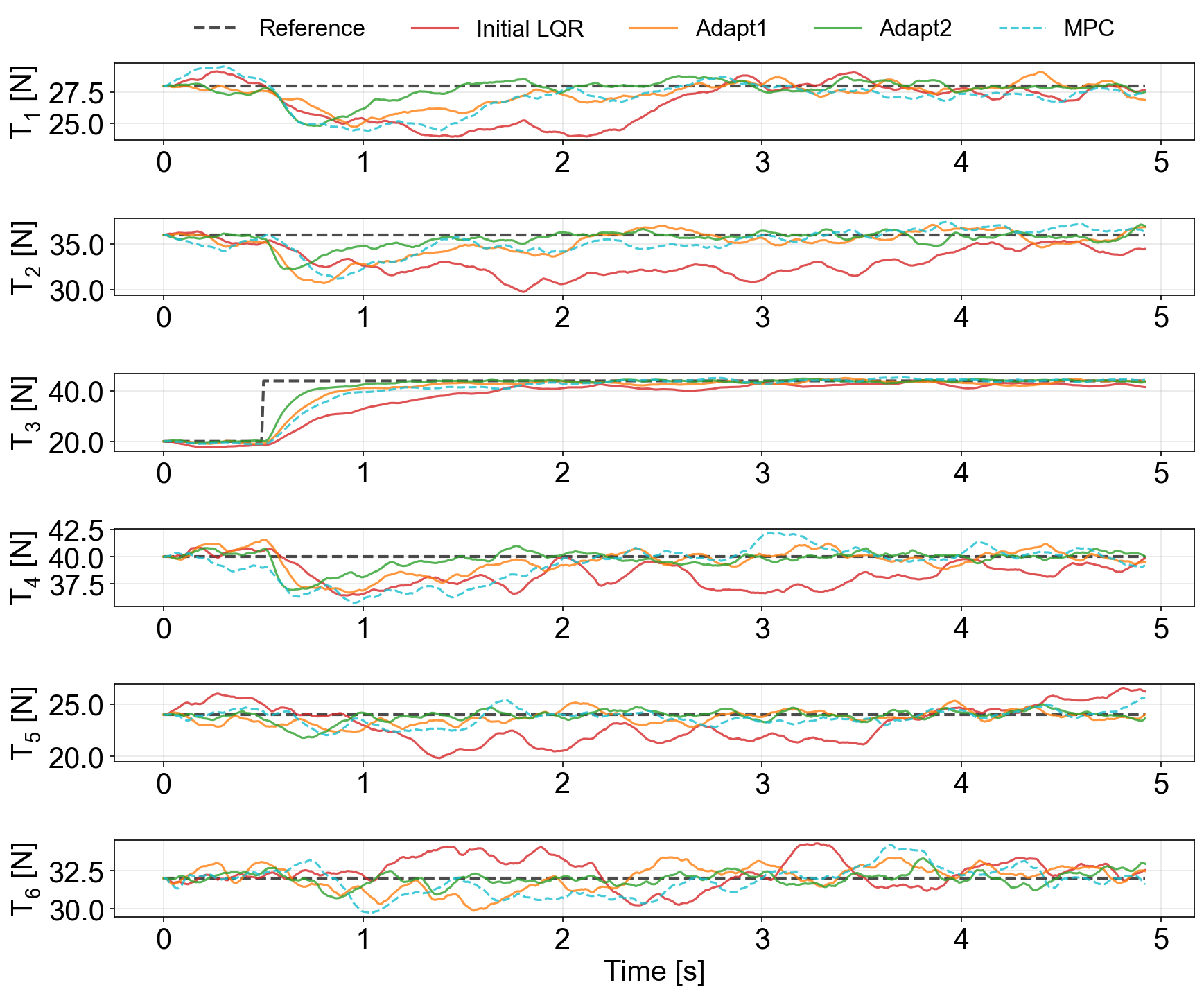}
    \caption{Comparative tension tracking performance across three adaptation cycles under identical conditions}
    \label{fig:tension_comparison}
\end{figure}

\paragraph{Tension Control Performance Under Identical Conditions}
Figure~\ref{fig:tension_comparison} compares controllers Initial--Adapt2 
and the MPC baseline under identical conditions to isolate performance 
improvements. Initial exhibits sustained errors exceeding 3~N in spans 
$T_1$, $T_2$, $T_4$, and $T_5$. The MPC baseline reduces these errors 
but still shows visible deviations from reference in several spans 
(RMSE: 1.6178~N). Adapt2 achieves tracking within $\pm 1$~N 
(RMSE: 1.0449~N), outperforming MPC by 35.4\%.

\subsection{Change in Velocity Setpoint}

\paragraph{System Configuration}
The velocity tracking scenario evaluates controller robustness during process acceleration transients common in R2R manufacturing startup and grade change operations. All six web spans maintain constant reference tensions at $T_i = 30$ N throughout the simulation. The unwinding velocity executes a step increase from $v_0 = 0.01$ m/s to $v_0 = 0.10$ m/s at $t = 1$ s, representing a 10-fold acceleration. Reference roller velocities for all actuated rollers are computed according to Equation~\eqref{eq:reference_velocity} to maintain equilibrium tension distribution during the velocity transient.

\begin{figure}[htbp]
    \centering
    \includegraphics[width=0.5\textwidth]{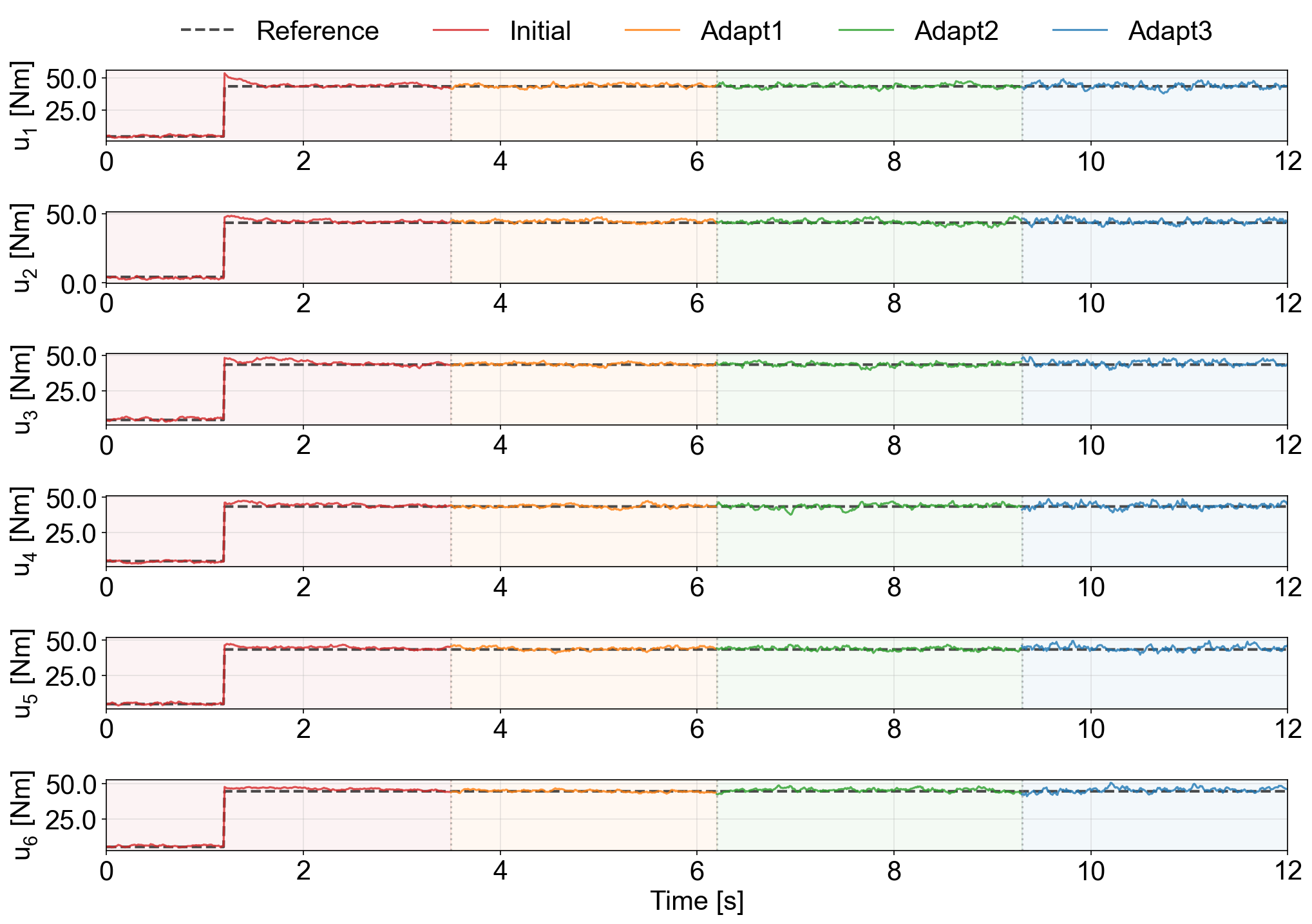}
    \caption{Control input trajectories during velocity setpoint change}
    \label{fig:velocity_sequential_control}
\end{figure}

\begin{figure}[htbp]
    \centering
    \includegraphics[width=0.5\textwidth]{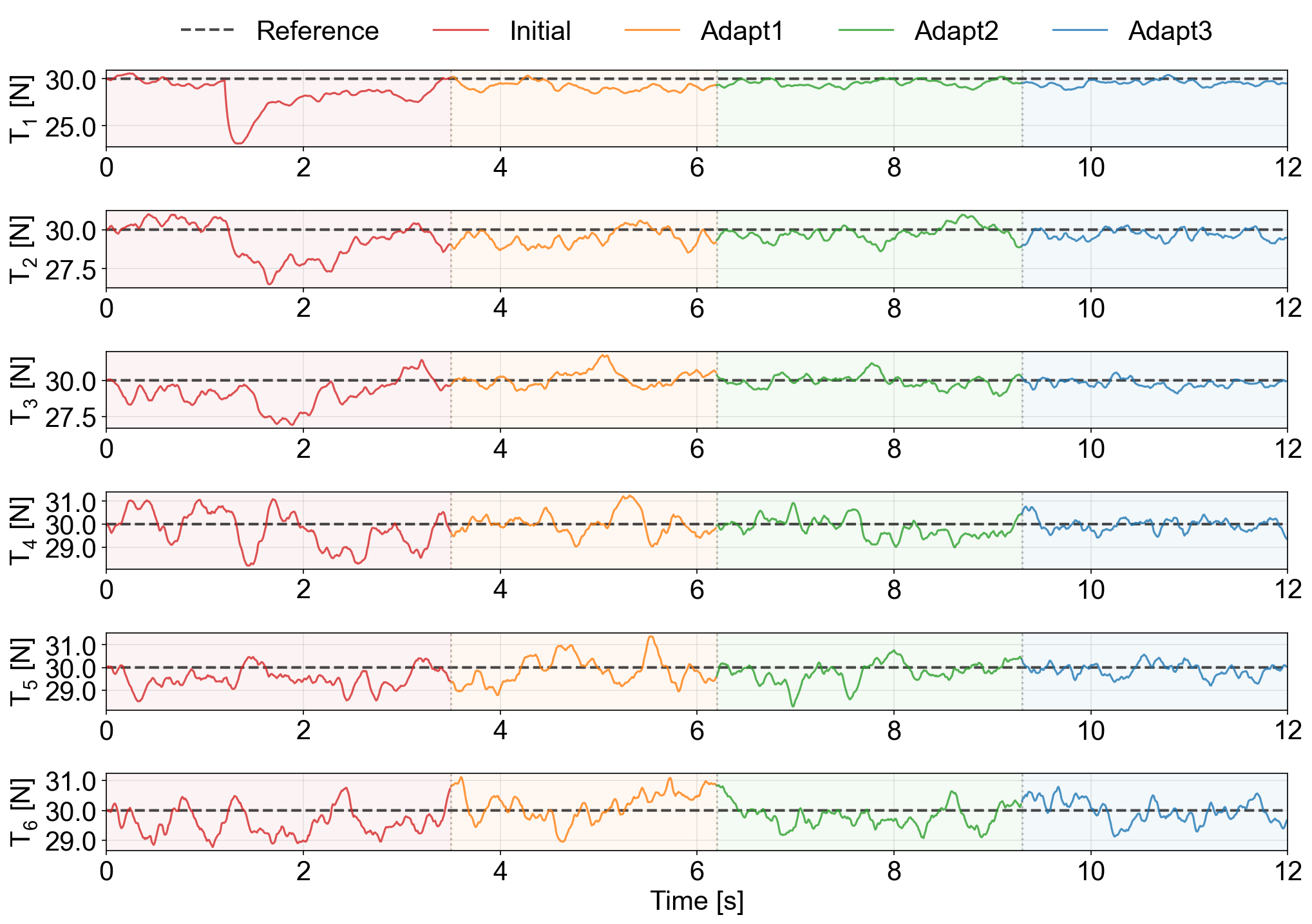}
    \caption{Tension trajectories during velocity setpoint change}
    \label{fig:velocity_sequential_tensions}
\end{figure}

\paragraph{Controller Adaptation}
The same LQR controller selected in the tension control scenario was deployed for velocity tracking validation. Figure~\ref{fig:velocity_sequential_control} and Figure~\ref{fig:velocity_sequential_tensions} present control inputs and tension trajectories during sequential adaptation. Initial exhibits poor tension regulation during and after the velocity ramp, with multiple spans showing oscillatory behavior and deviations from the 30 N setpoint. The Adaptation Agent identified insufficient feedforward compensation during velocity transients and excessive feedback gains causing post-transient oscillations. The suggested parameter adjustments, validated through simulation safety filter, were deployed as Adapt1, demonstrating reduced oscillations but persistent steady-state errors in several spans. Adapt2 and Adapt3 achieve improved regulation with tensions remaining closer to reference during the acceleration phase and subsequent steady-state operation.


\begin{figure}[htbp]
    \centering
    \includegraphics[width=0.42\textwidth]{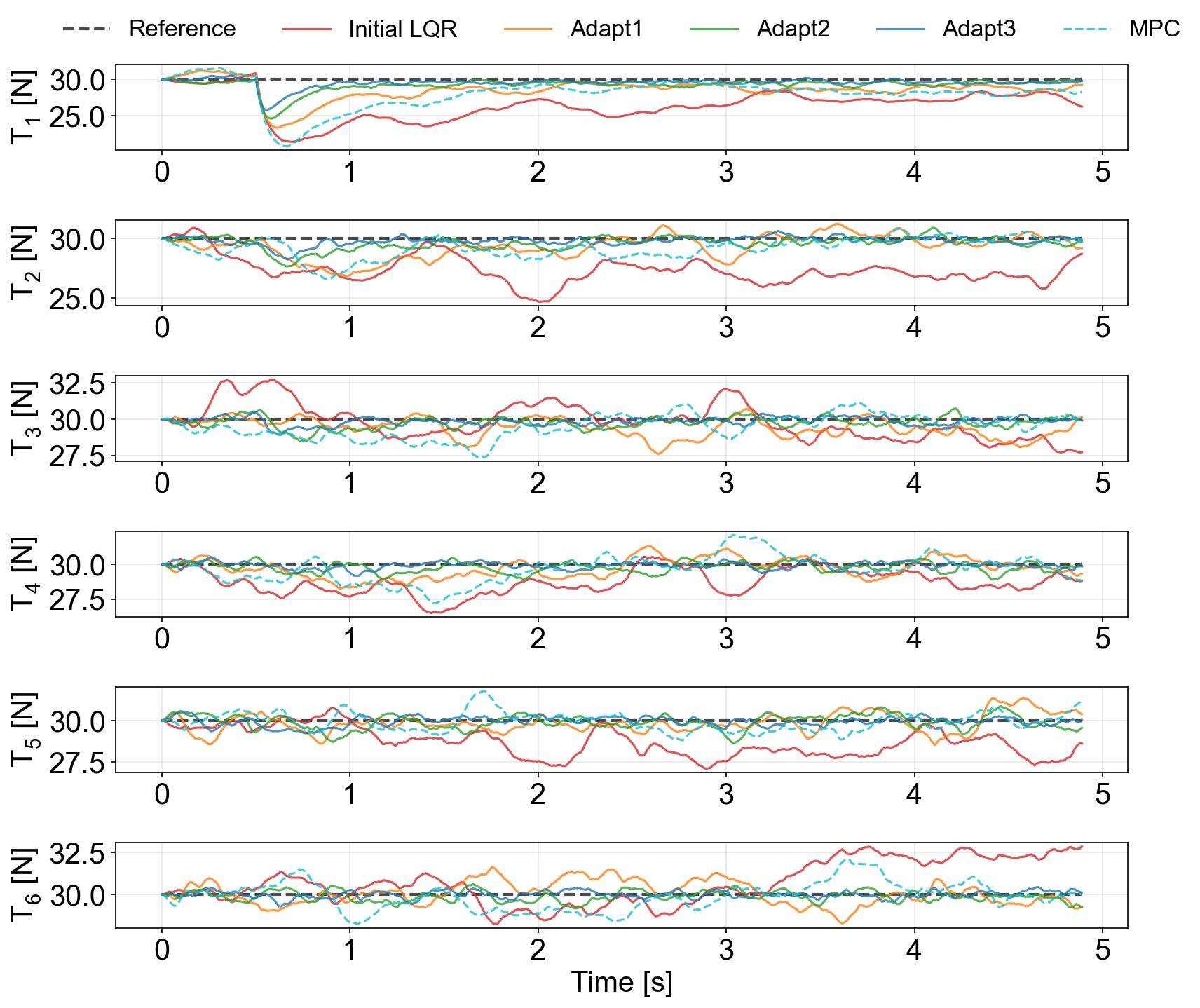}
    \caption{Comparative tension regulation performance during velocity ramping across four adaptation cycles under identical conditions}
    \label{fig:velocity_comparison}
\end{figure}

\paragraph{Velocity Control Performance Under Identical Conditions}
Figure~\ref{fig:velocity_comparison} compares the initial controller 
Initial with three sequentially adapted versions (Adapt1, Adapt2, Adapt3) 
and the MPC baseline under identical velocity ramping conditions. Initial 
shows sustained tension deviations exceeding 4~N in spans $T_1$, $T_2$, 
$T_3$, and $T_5$ during both transient and steady-state phases. The MPC 
baseline performs better than Initial but still shows 
deviations (RMSE: 1.0102~N). Adapt1 reduces transient deviations substantially, while Adapt2 and Adapt3 maintain tensions within $\pm 1$~N throughout the acceleration and post-transient periods, outperforming MPC by 70.9\%.

\subsection{Performance Summary}

Table~\ref{tab:performance_summary} summarizes the RMSE performance across both validation scenarios. The LLM-assisted adaptation achieves consistent improvement over iterations, with final performance (Adapt2/Adapt3) outperforming both the initial deployment and the MPC baseline in both scenarios.

\begin{table}[htbp]
\centering
\caption{RMSE Performance Comparison Across Validation Scenarios}
\label{tab:performance_summary}
\begin{tabular}{lcc}
\hline
\textbf{Configuration} & \textbf{TensionControl (N)} & \textbf{VelocityTrack (N)} \\
\hline
Initial & 2.3615 & 1.6708 \\
Adapt1 & 1.4267 & 0.7902 \\
Adapt2 & 1.0449 & 0.4490 \\
Adapt3 & --- & 0.2944 \\
\hline
MPC (Baseline) & 1.6178 & 1.0102 \\
\hline
\end{tabular}
\end{table}

\section{Conclusion and Future Work}

This work presents an LLM-assisted multi-agent control framework that integrates LLM with controller design for R2R manufacturing systems. The framework explores sim-to-real adaptation through simulation-validated tuning, working toward stable tension control and velocity tracking despite major model mismatch. Validation studies shows controller selection, iterative adaptation, and monitoring with diagnostic capabilities. Future work will focus on: (1) hardware validation on physical R2R testbeds; (2) extension to other manufacturing processes beyond R2R systems; and (3) reuse of learned controller configurations and adaptation histories from one R2R line to accelerate commissioning on new lines with different materials or geometries.

\section*{Acknowledgements}
This work is based upon work partially supported by the National Science Foundation under Cooperative Agreement No. CMMI-2041470. Any opinions, findings and conclusions expressed in this material are those of the author(s) and do not
necessarily reflect the views of the National Science Foundation. 

\appendix

\appendix

\section{Detailed Prompt for Each Agent}
\label{sec:agent_prompts}

This section provides the complete system prompts that define the role, inputs, and expected outputs for each specialized agent in the framework.

\subsection{SysID Agent System Prompt}

\begin{mdframed}[linewidth=1pt, linecolor=gray, backgroundcolor=gray!10]
\textbf{Agent Role:} System Identification Expert

\textbf{Your task:} Construct a  physics-informed simulation model of the R2R manufacturing system through systematic parameter identification.

\textbf{Inputs provided:} (1) Historical operational data containing control inputs $u_k$, measured outputs $y_k$ (tensions and velocities), and timestamps $t_k$; (2) System dynamic equations (Equations \eqref{eq:web_tension_dynamics} and \eqref{eq:roller_velocity_dynamics}); (3) RAG access to domain knowledge on R2R system identification best practices; (4) Physical parameter bounds and constraints.

\textbf{Required outputs:} (1) Identified parameter vector $\theta$ with confidence intervals; (2) Validation metrics ($R^2$ scores for tension and velocity predictions, mean absolute errors); (3) Constructed simulation model $\mathcal{S}_{sim}$ stored in System Model Repository; (4) Assessment of model fidelity and recommendations for controller design phase.
\end{mdframed}

\subsection{Initial Control Agent System Prompt}

\begin{mdframed}[linewidth=1pt, linecolor=gray, backgroundcolor=gray!10]
\textbf{Agent Role:} Controller Design and Selection Expert

\textbf{Your task:} Design, compare, and select the optimal controller architecture for the R2R tension control system through systematic simulation-based evaluation.

\textbf{Inputs provided:} (1) Validated simulation model $\mathcal{S}_{sim}$ from Phase 0; (2) Control objectives and constraints (Equations \eqref{eq:control_objective}, \eqref{eq:constraints}); (3) Three candidate controller architectures (PID, MPC, LQR) with their mathematical formulations (Equations \eqref{eq:pid}, \eqref{eq:mpc}, \eqref{eq:lqr}); (4) Performance metrics for evaluation (Equations \eqref{eq:rmse} through \eqref{eq:robustness}); (5) Code Agent for implementation and testing.

\textbf{Required outputs:} (1) Tuned hyperparameters for each controller architecture following Algorithm \ref{alg:hyperparameter_tuning}; (2) Comparative performance table with all metrics; (3) Selection decision with detailed engineering justification; (4) Optimal controller $C^*$ with configuration stored in Controller Configuration Storage for Phase 2 deployment.
\mbox{}\end{mdframed}

\subsection{Adaptation Agent System Prompt}

\begin{mdframed}[linewidth=1pt, linecolor=gray, backgroundcolor=gray!10]
\textbf{Agent Role:} Sim-to-Real Adaptation Specialist

\textbf{Your task:} Perform rapid sim-to-real controller adaptation to address model mismatch between simulation and physical hardware through iterative tuning validated by simulation safety filter.

\textbf{Inputs provided:} (1) Optimal controller $C^*$ with parameters $\phi^*$ from Phase 1; (2) Real system response data from deployment tests; (3) Simulation model $\mathcal{S}_{sim}$ for safety validation; (4) Performance convergence criteria (Equation \eqref{eq:convergence_criteria}); (5) Code Agent for testing adjustments.

\textbf{Required outputs:} (1) Diagnostic analysis identifying specific model mismatch symptoms and root causes; (2) Suggested parameter adjustments $\Delta\phi$ with control-theoretic justification; (3) Simulation safety filter validation results per Equation \eqref{eq:safety_approval}; (4) Adapted controller parameters $\phi_{final}$ when convergence criteria satisfied; (5) Complete adaptation history stored in Performance Baseline Database.

\textbf{Design process:} Follow Algorithm \ref{alg:sim_to_real} for structured iteration.
\end{mdframed}

\subsection{Monitoring Agent System Prompt}

\begin{mdframed}[linewidth=1pt, linecolor=gray, backgroundcolor=gray!10]
\textbf{Agent Role:} Performance Monitoring and Diagnostics Expert

\textbf{Your task:} Perform continuous dual-layer performance monitoring and diagnostic analysis during stable production operation to detect degradation and distinguish between control-adjustable issues and physical maintenance requirements.

\textbf{Inputs provided:} (1) Real-time operational data stream $\mathcal{D}_{live}$ containing tensions, velocities, and control signals; (2) Baseline performance vector $P_{baseline}$ established at Phase 2 completion; (3) Degradation detection threshold per Equation \eqref{eq:degradation_threshold}; (4) Simulation model $\mathcal{S}_{sim}$ for validating adaptive adjustments; (5) Code Agent for implementing diagnostic algorithms.

\textbf{Required outputs:} (1) \textbf{Layer 1 analysis}—Continuous performance metric evaluation $\Delta P(t)$ with degradation flags; (2) \textbf{Layer 2 analysis}—Root cause diagnostics when degradation detected, categorizing issues as material property variations, mechanical degradation, sensor problems, environmental factors, or process disturbances; (3) Diagnostic reports with confidence-weighted hypotheses distinguishing control vs. maintenance issues; (4) For control-related issues: suggested parameter adjustments validated through simulation safety filter; (5) For physical issues: maintenance alerts with detailed evidence and recommended actions; (6) Adaptation history logged in Diagnostic History Log.

\textbf{Design process:} Follow Algorithm \ref{alg:adaptive_adjustment} for structured response protocol.
\end{mdframed}

\end{document}